\documentclass[twoside,12pt]{article}
\usepackage{epsfig}

\def\Journal#1#2#3#4{{#1} {#2} (#4) #3 }

\def\NPA{{\em Nucl. Phys.} A}
\def\NPB{{\em Nucl. Phys.} B}

\def\NPB{{\em Nucl. Phys.} B}

\def\PLB{{\em Phys. Lett.} B}

\def\PRL{\em Phys. Rev. Lett.}

\def\PREP{\em Phys. Rep.}

\def\PRD{{\em Phys. Rev.} D}
\def\PRC{{\em Phys. Rev.} C}

\newcommand{\be}{\begin{equation}}
\newcommand{\ee}{\end{equation}}
\newcommand{\bea}{\begin{eqnarray}}
\newcommand{\eea}{\end{eqnarray}}

\begin{document}

\title{ \vspace{1cm} Neutrinos in Cosmology and Astrophysics}
\author{A.B. \ Balantekin,$^{1}$\\ 
G.M.\ Fuller$^2$\\ 
\\
$^1$Physics Department, University of Wisconsin, Madison WI 53706 USA\\
$^2$Department of Physics, University of California, San Diego, La Jolla CA 92093 USA\\
}
\maketitle
\begin{abstract} 
We briefly review the recent developments in neutrino physics and astrophysics which have import
for frontline research in nuclear physics. These developments, we argue, tie nuclear physics
to exciting developments in observational cosmology and astrophysics in new ways. Moreover,
the behavior of neutrinos in dense matter is itself a fundamental problem in many-body
quantum mechanics, in some ways akin to well-known issues in nuclear matter and nuclei, and in some ways radically different, 
especially because of nonlinearity and quantum de-coherence. The self-interacting neutrino gas is the only many body system driven by the weak interactions. 
 
\end{abstract}
\section{Introduction}

Experiments have now revealed many of the fundamental properties of neutrinos, including their mass-squared differences and three (mixing angles $\theta_{1 2}$, $\theta_{2 3}$, $\theta_{1 3}$) of the four parameters characterizing the unitary transformation between the neutrino vacuum energy (mass) states and the weak interaction (flavor) eigenstates (a recent review is given in 
\cite{Balantekin:2013tqa}). We now lack only the fourth parameter, the CP-violating phase, and the absolute neutrino rest masses and the way these are arranged, {\it i.e.,} whether nature has chosen the normal or inverted mass hierarchy. 

Moreover, the fact of nonzero neutrino rest masses immediately begs the question of whether there exist right-handed, so-called sterile neutrinos. There are many models for sterile neutrino states which are not really sterile by virtue of vacuum mixing with ordinary, active neutrinos. The mass scales of these sterile species are not well predicted in these models, and can range from masses comparable to the unification scale, all the way down to the sub-eV regime. There are even intriguing experimental hints for the existence of sterile neutrinos with masses in the eV range.

All of these neutrino properties, measured and unmeasured, may figure prominently in key astrophysical environments, for example in core collapse supernovae and in the early universe and Big Bang Nucleosynthesis (BBN). This connects fundamental neutrino physics to the breathtaking advances in astrophysical numerical modeling and to the near-fantastic increase in the volume and scope of observational data obtained from both from ground- and space-based observatories. The reasons for the neutrino-astrophysics tie-in are twofold: neutrinos can carry and transport the bulk of the entropy and energy in these environments, along the way influencing composition; and the weak interaction, including neutrino interactions, is unique in being able to change isospin, {\it i.e.,} inter-converting neutrons and protons. The latter issue is fundamentally dependent on the flavor states of the neutrinos, and these can change.

The charge to neutrino physicists and astrophysicists is then clear: we must be able to calculate how neutrinos change their flavors as they move from dense nuclear matter-like environments (either in the Early Universe or in supernova proto-neutron star cores) to relatively low density environments (like the supernova envelope or the post-weak decoupling epoch in the early universe). Historically this neutrino flavor evolution problem in astrophysics has been approached with a \lq\lq separation of scales.\rq\rq\ 

At relatively low densities a
Schr\"odinger-like equation governs the coherent limit, where forward scattering of neutrinos on electrons, quarks/nucleons, and other neutrinos dominates over inelastic and direction-changing processes, and in-medium oscillation lengths are short compared to neutrino mean free paths. At high densities, a Boltzmann treatment of neutrino energy, number, and heat transport is used. In that limit neutrino inelastic and direction-changing scattering is dominant and flavor oscillations are ignored because oscillation lengths are long compared to mean free paths. The way the problem has been approached in supernova environments, the Boltzmann equation is used in the neutron star and in the region immediately above it, and the Schr\"odinger/coherent approach is employed further out in the envelope where the density is lower. We now know that this separation of scales fails in some cases, an outstanding example of which is the \lq\lq neutrino halo\rq\rq\ effect \cite{Cherry:2012zw}. 

To follow neutrino flavor evolution in the general case, i.e. in a medium of any density, requires a set of full neutrino quantum kinetic equations (QKE's). Obviously, these equations should reduce to: (1) the Boltzmann transport equation in the high density limit where scattering-induced de-coherence dominates and flavor conversion can be neglected; and (b) a Schr\"odinger-like equation in the low density limit. However, all manner of plausible-looking QKE's have the same asymptotic limits, and this has necessitated Vlasenko, Fuller, and Cirigliano \cite{VFC} 
 to derive them from fundamental considerations in quantum field theory. This produces QKE's broadly similar to those found in Raffelt \& Sigl \cite{Sigl:1992fn} and Strack \& Burrows \cite{Strack:2005ux}. Traditionally the neutrino flavor evolution problem in astrophysics has been approached with a \lq\lq separation of scales.\rq\rq\

Solving the QKE's is a numerical nightmare, essentially differing from traditional neutrino transport calculations through (sometimes very) high frequency quantum phases. Hence the appeal of a separation of scales approach is clear. But as we shall see below, even that approach, flawed as it is, is fraught with unresolved many-body physics problems.

\section{Cosmology}

Neutrinos influence almost every aspect of the physics of the early universe. This fact, combined with several recent, or expected future, developments in observational cosmology and neutrino experiment promise to \lq\lq box-in\rq\rq\ any new physics in the neutrino sector. There are five key developments.

1) Observations of the cosmic microwave background (CMB) radiation have given us a precise measurement of the baryon content of the universe: this corresponds to a baryon-to-photon ratio $\eta \approx 6.11\times{10}^{-10}$. Future observations will get this number to better than $1\%$ precision. This quantity is a key parameter in Big Bang Nucleosynthesis (BBN). Moreover, global minimization of the CMB and other cosmological data gives us the cosmological parameters, the age and the curvature parameter.

2) The advent of large, ten-meter class telescopes like Keck, has revealed the primordial deuterium abundance, again to fairly good precision 
\cite{Kirkman:2003uv}. This is significant, because the chief determinant of the $^2{\rm H}$ yield in BBN is the baryon density, which we know very well from 1). Any discrepancy between the observed primordial deuterium and that predicted by BBN calculations using the CMB-determined baryon density could have its origin in neutrino sector physics. Though the dependence of the BBN deuterium yield in BBN is much weaker than that of $^4{\rm He}$, we may be able to infer the deuterium primordial abundance with more confidence and fewer systematic issues.

3) Observations of the Silk damping tail (higher wave number end) of the CMB power spectrum can give a measure of the primordial $^4{\rm He}$ abundance. This determination is completely independent of the value obtained via the linear regression/compact blue galaxy approach. A high-precision determination of primordial helium abundance combined with 1) and 2), is highly restrictive of new physics in the neutrino sector.

4) CMB observations can give a measure of the ratio of energy density in relativistic particles to that carried by particles with nonrelativistic kinematics at the epoch of photon decoupling. This epoch corresponds to photon temperature $T_{\rm CMB} \approx 0.2\,{\rm eV}$. By convention, the relativistic energy density $\rho_{\rm rad}$ at this epoch is expressed in terms of a parameter $N_{\rm eff}$,
\begin{equation}
\rho_{\rm rad} = {\left[  2+\frac{7}{4} {\left( {\frac{4}{11}} \right)}^{4/3}\, N_{\rm eff}  \right]}\,
{\frac{\pi^2}{30}}\,T_{\rm CMB}^4 .
\label{Neff}
\end{equation}
With this definition, standard model physics and cosmology predicts $N_{\rm eff} \approx 3.046$  \cite{Dicus:1982bz}. 
The excess over $3$, 
corresponding to three flavors of neutrinos 
with black body, Fermi-Dirac-shaped energy spectra, arises from 
 $e^\pm$-pair annihilation into out of equilibrium neutrino pairs near and during the BBN epoch. 
It is important to note that $N_{\rm eff}$ parameterizes {\it all} relativistic energy density 
at the photon decoupling epoch, not just that contributed specifically by the known active neutrinos. 
Any measurement of $N_{\rm eff}$ significantly different from $3.046$, either lower or higher, 
signals new physics, either new particle physics, or some deviation in the 
history of the early universe from that predicted by the standard model. Current CMB measurements of $N_{\rm eff}$ are consistent with the standard model, but have large uncertainties. In the near future the Planck satellite collaboration will report an analysis of their data which should give $N_{\rm eff}$ to $10\%$ \cite{planck}.

5) Finally, the CMB plus observations of smaller-scale large scale structure, {\it e.g.,} the Lyman alpha forest, promise a good limit on what is usually termed the sum of the light neutrino masses, $\sum m_\nu$. The best constraints in this regard will probably come from experiments that utilize weak gravitational lensing of the CMB, and these may well get down to the $\sum{ m_\nu} < 0.1\,{\rm eV}$ range which could in principle see a signal for neutrino mass if nature has chosen the inverted neutrino mass hierarchy \cite{Abazajian:2011dt}. 
 This is possible even with no neutrino mass degeneracy offset. It should be noted, however, that these observations do not actually measure the sum of the vacuum neutrino rest mass eigenvalues, but rather effectively a convolution of these with the energy spectra of the constituent neutrinos. As such, measurement of, or constraints on, $\sum{ m_\nu}$ are tantamount to a probe of the relic neutrino energy spectrum and density, once the rest masses are known.

Sterile neutrinos are a case in point when it comes to the constraining or revealing power of the observations/considerations in 1)-5). The experimental and observational constraints on sterile neutrinos are discussed in Refs.~ \cite{Kusenko:2009up} and \cite{Abazajian:2012ys}. 
For example, if there were a sterile neutrino with a rest mass $\sim 1\,{\rm eV}$, that mixed in vacuum with active neutrino species at the rather large level suggested by the mini-BooNE experiment or the reactor neutrino anomaly, we might expect someday to see a significant impact on measurements in items 2), 3), 4), and 5); or some subset of these which would be revealing of additional new physics, like dilution from particle decay
\cite{Fuller:2011qy}. 
Calculating just how active-sterile neutrino mixing in a case like this might distort $\nu_e$ and $\bar\nu_e$ energy spectra at the BBN epoch could be a tricky issue, however, necessitating a QKE solution.

\section{Core-collapse supernovae}

Core-collapse supernovae, like some epochs in the early universe, are neutrino-dominated dynamical systems. 
In these supernovae essentially all the gravitational energy released during collapse escapes in the form of intense neutrino fluxes emerging from the newly-born neutron star. During the first ten seconds or so of the existence of the neutron star, these neutrino fluxes drive a wind from its surface in which various nuclear species may be synthesized. 
In a core-collapse supernova environment neutrino-neutrino interactions are not negligible, as the gravitational binding energy of the progenitor massive star is converted into $\sim 10^{58}$ neutrinos during the cooling process of the proto-neutron star.  The total energy carried by those neutrinos is 10$^{53}$ ergs, as compared to the total optical and kinetic energy of these events which is 10$^{51}$ ergs. 
The interactions between those copious neutrinos lead to novel collective and emergent effects, such as conserved quantities and interesting features in the neutrino energy spectra. Collective neutrino oscillations play a crucial role both for neutrinos and antineutrinos. There is a growing literature on the collective neutrino oscillations, a good starting point is a recent review  
\cite{Duan:2010bg}. Collective neutrino oscillations produce an interesting effect, called spectral swappings or splits, on the final neutrino energy spectra: at a particular energy these spectra are almost completely divided into parts of different flavors \cite{Raffelt:2007cb,Duan:2008za}. 

It is interesting to note that core-collapse supernovae are the only many-body systems driven by the weak interactions (see Table \ref{tab:many}). 
\begin{table}
\begin{center}
\begin{minipage}[t]{16.5 cm}
\caption{Many-body systems in physics.}
\label{tab:many}
\end{minipage}
\begin{tabular}{c|c|c}
\hline
\textbf{System} & \textbf{Primary interaction} & \textbf{Number of particles}\\
\hline
Nuclei&  Strong & at most $\sim$250 particles \\
\hline 
Condensed matter& Electromagnetic & at most $N_A$ particles\\
\hline
Neutrinos in SN & Weak& $\sim 10^{58}$particles \\
\hline
\end{tabular}
\end{center}
\end{table}
This table nicely illustrates that astrophysical extremes allow us to test physics that cannot be tested elsewhere: Neutrino-neutrino interactions,  which represent a part of the Standard Model, are not accessible with any other experimental tools. 

A complete theoretical treatment of all the many-body effects due to neutrino-neutrino interactions would be very complicated and usually several simplifying assumptions are made. The coherent scattering of the neutrinos off one another is considered dominant. Even with this restriction solving the full many-body problem is exceedingly difficult. Instead a mean-field approximation which represents the saddle-point solution of the path integral for the full many-body system  \cite{Balantekin:2006tg} is typically used. In addition, the Hamiltonian describing the system depends on the angles between all the pairs of neutrino momenta. Earlier calculations employed an average of these angles ("single-angle" approximation), however increasingly sophisticated multi-angle calculations are now available. A recent calculation with three flavors finds that  multi-angle formulation reduces the adiabaticity of flavor evolution in the normal neutrino mass hierarchy, resulting in lower swap energies \cite{Cherry:2010yc}. It seems that the single-angle approximation seems to be sufficient in some cases, but is inadequate in other situations.  

The saddle-point approximation effectively reduces the full neutrino Hamiltonian with one- and two-body terms to an one-body Hamiltonian. 
This is reminiscent of the random-phase approximation in many-body theory where quadratic products of the operators are "linearized" by replacing one of them with a "mean-field" value.  
Corrections to the saddle-point approximation are expected to be small, but they have not yet been calculated. In the single-angle limit, using a formal analogy between the many-neutrino Hamiltonian and the Hamiltonian describing  BCS superconductivity, 
one can write down the conserved quantities of the system \cite{Pehlivan:2011hp}. It turns out that the invariants of the full Hamiltonian are also invariants of the one-body Hamiltonian when they are properly linearized.This provides further confidence in the aptness of the linearization procedure itself. 

Another assumption which was recently relaxed is the assumption of forward scattering. Neutrinos that scatter in non-forward directions could create a "neutrino halo"  that would interact with the other outgoing neutrinos. The fraction of outflowing neutrinos interacting with this neutrino halo is significant  \cite{Cherry:2012zw}. The halo could be a significant effect in every supernova environment except very late time neutrino driven wind. It was argued that 
the multiangle effects could suppress self-induced flavor conversion during the accretion phase \cite{Sarikas:2012vb}. However, the halo changes the nature of the flavor evolution, turning it into a boundary value problem instead of an initial value one. 
A full numerical treatment of the halo, taking into account this effect, has been only performed for O-Ne-Mg core-collapse supernovae \cite{Cherry:2013mv}.

Core-collapse supernovae are likely sites for several nucleosynthesis scenarios. One of these is nucleosynthesis via neutrino-induced nucleon emission (the $\nu$-process) \cite{Woosley:1989bd}. For example, the conversion of $^{20}$Ne into $^{19}$F in the outer shells via neutrino capture would account for the entire observed abundance of $^{19}$F. In the absence of collective oscillations, one expects a  
hierarchy $E_{\nu_e} < E_{\bar{\nu}_e} < E_{\nu_{\mu},\nu_{\tau},{\bar{\nu}}_{\mu},{\bar{\nu}}_{\tau}}$ in the energy spectra of the neutrino fluxes that pass through those outer shells. While the MSW resonance governed by $\delta m_{21}^2$ is at solar densities, the resonance governed by  $\delta m_{31}^2$ is at matter densities that exist in those outer shells of a supernova. It was recently pointed out that 
MSW effect for the inverted hierarchy, by converting the more energetic muon and tau antineutrinos into electron antineutrinos, boosts the 
$\nu$-process nucleosynthesis yields of $^{11}$B and $^7$Li \cite{Yoshida:2005uy}. In the normal hierarchy this would not happen: it is interesting to be able to relate the elemental abundances to the neutrino hierarchy. One caveat is that  once the neutrinos reach the He shells, complete swappings between electron neutrinos (or antineutrinos) and other flavors due to the collective neutrino oscillations would not be distinguishable from the adiabatic MSW oscillations \cite{Fogli:2007bk}. 

The site of the r-process nucleosynthesis is an open question \cite{Qian:2007vq}. One needs a site which is the isospin mirror of the Early Universe, a hot gas expanding and condensing into nuclei as it cools. The high-temperature, high-entropy region outside the newly-formed neutron star in a core-collapse supernova was suggested to be an r-process site \cite{Meyer:1992zz}. 
The neutrino-driven wind, one candidate site where the r-process may take place, yields about the observed amount of the r-process nuclei. Current hydrodynamical simulations of the neutrino-driven wind do not seem to  reach the extreme conditions necessary for the r-process \cite{wind}. Since collective neutrino oscillations dominate the neutrino propagation much deeper than the conventional matter-induced MSW effect, they would also impact r-process nucleosynthesis yields if the neutrino-driven winds are shown to be the appropriate sites \cite{Balantekin:2004ug,Duan:2010af}. There are other suggested sites for the r-process nucleosynthesis. They include $^4$He mantles of the metal-poor (i.e., early) supernova progenitors \cite{Banerjee:2011zm} and neutron-star mergers \cite{Argast:2003he}.  

Electron fraction, or equivalently neutron-to-proton ratio (a controlling parameter for nucleosynthesis) is determined by the neutrino capture rates: 
\be 
\nu_e + n \rightleftharpoons  p + e^-
\ee
and 
\be
\bar{\nu}_e + p \rightleftharpoons n + e^+, 
\ee
Hence, aside from driving the wind, the most important impact of the neutrino fluxes for a potential r-process is that neutrino interactions on free nucleons set the neutron richness of the outflow. Neutrino-nucleus interactions can also leave a noticeable imprint on the distribution of synthesized nuclei \cite{Balantekin:2003ip}.  A summary of neutrino processes relevant for flavor evolution in core-collapse supernovae is given in Fig. 1. 

\begin{figure}
\begin{center}
\includegraphics[width=17cm]{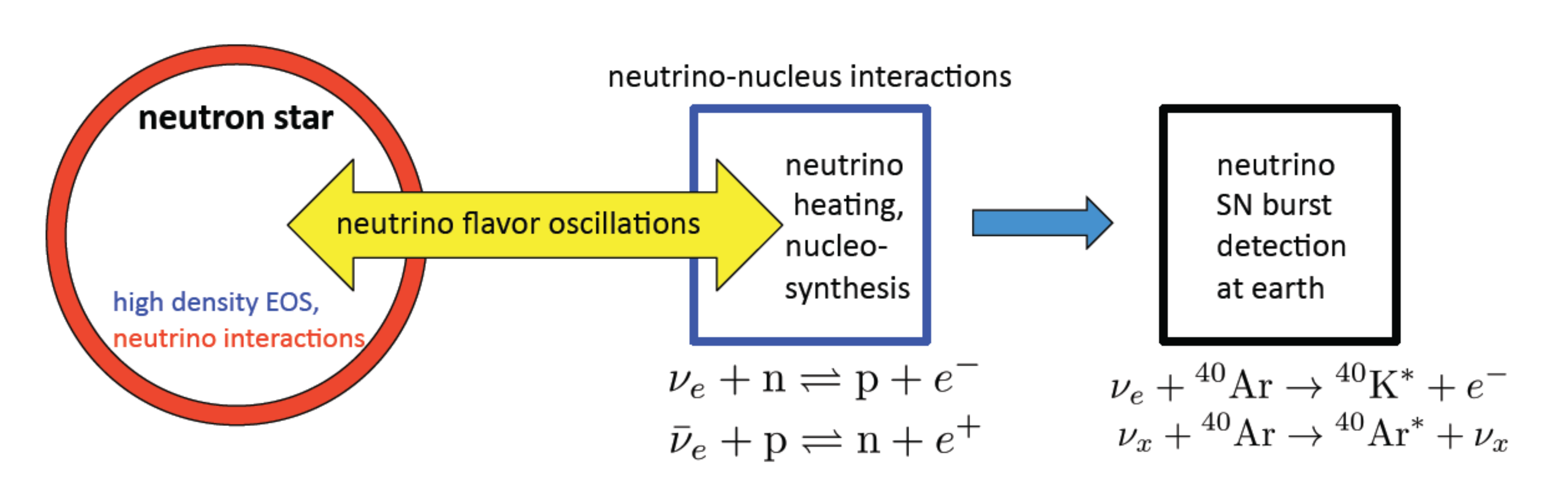}
\caption{A summary of neutrino processes in core-collapse supernovae, highlighting the importance of neutrino flavor evolution.}
\label{fig:1}
\end{center}
\end{figure}

Progress in not only in calculating r-process nucleosynthesis but also in a number of research frontiers in nuclear astrophysics depends on understanding spin-isospin response in a broad range of nuclei from stable isotopes to rare ions that can be studied in dedicated facilities. 
Currently many major accelerator projects around the world, at different stages of construction and operation, aim to explore the physics of these exotic rare nuclei. Neutrinos indeed bridge the cutting-edge experimental efforts at the rare isotope facilities and laboratory probes of spin-isospin response of nuclei with nuclear astrophysics efforts aimed at learning about the origin of elements.

%

\section{Conclusion}
 
It can be argued that neutrino rest mass and vacuum flavor mixing is physics beyond the Standard Model. Certainly the existence of sterile neutrino states falls into this category. Since neutrinos carry a dominant fraction of the total energy and even entropy in the Early Universe, core collapse supernovae and compact object merger environments, and since these venues can be the sites of key nucleosynthesis events, their dynamics and local interactions may be important to understand. Furthermore, since the most important neutrino interactions for nucleosynthesis, the charged current isospin-changing reactions, are flavor dependent, this understanding will come only when neutrino flavor transformation in medium is understood. 

As a consequence, we believe that the nonlinear neutrino flavor transformation problem may lie at the heart of many important problems in nuclear physics and astrophysics. These problems include the origin of the lightest and heaviest nuclei in the nuclear astrophysics realm. On the pure nuclear physics side, the many-body techniques required to solve the neutrino transport and flavor evolution problems echo the techniques and insights developed to understand nuclear matter and nuclear structure.

Given the expected golden future for observational cosmology, the real if chancy possibility of catching a Galactic core collapse supernova neutrino burst in a new generation of terrestrial detectors, and the anticipated future detection of compact object mergers with Advanced LIGO, we believe that a deeper understanding of neutrino flavor dynamics should be a goal for some of us in the nuclear theory community. 

This work was supported in part 
by the U.S. National Science Foundation Grants No.  PHY-1205024 (U. Wisconsin)
and PHY-0970064 (U. California, San Diego), in part by the University of Wisconsin Research Committee with funds
granted by the Wisconsin Alumni Research Foundation, in part by the University of California Office of the President, and in part by the LANL/DOE topical collaboration.

\end{document}